\documentclass[%
 reprint,
superscriptaddress,
 amsmath,amssymb,
 aps,
]{revtex4-2}
\usepackage{xcolor}
\usepackage{hyperref}

\usepackage{graphicx}% Include figure files
\usepackage{dcolumn}% Align table columns on decimal point
\usepackage{bm}% bold math
\usepackage{array}
\usepackage{braket}
\usepackage{comment}
\excludecomment{hidden} 
\begin{document}

\preprint{APS/123-QED}

\title{Floquet Interpretation of Avoided Crossings \\ in AC Stark-Shifted Rydberg-EIT Spectra}% Force line breaks with \\

\thanks{christopher.holloway@nist.gov}%

\author{Rajavardhan~Talashila}
\affiliation{National Institute of Standards and Technology, Boulder, Colorado 80305, USA}
\affiliation{Department of Electrical, Computer, and Engineering, University of Colorado, Boulder, Colorado 80309, USA}
\author{Nikunjkumar~Prajapati} 
\affiliation{National Institute of Standards and Technology, Boulder, Colorado 80305, USA}
\author{Noah~Schlossberger}
\affiliation{National Institute of Standards and Technology, Boulder, Colorado 80305, USA}
\author{Christopher~L.~Holloway}
\affiliation{National Institute of Standards and Technology, Boulder, Colorado 80305, USA}
\email[]{christopher.holloway@nist.gov}

\date{\today}% It is always \today, today,

\begin{abstract}
We present a combined experimental and Floquet-theoretical study of avoided crossings in the AC Stark-shifted electromagnetically induced transparency (EIT) spectra of Rydberg atoms. Observed avoided-crossing structures cannot be fully explained by conventional AC Stark maps alone, which provide the energy shifts but do not reveal the underlying state composition and coupling pathways. Using the Shirley method, we analyze the Floquet eigenstates to establish a direct connection between the measured spectra and the underlying dressed-state dynamics. By tracking the evolution of state mixing, bare-state composition, and dominant higher-order Floquet coupling pathways, we reveal the physical mechanisms responsible for the formation of avoided crossings across different principal quantum numbers and RF frequencies. Our results demonstrate that Floquet-eigenstate analysis provides a powerful framework for interpreting complex AC Stark-shifted Rydberg-EIT spectra and the underlying higher-order Floquet coupling interactions.
\end{abstract}

%We present a combined experimental and Floquet-theoretical study of avoided crossings in the AC Stark-shifted electromagnetically induced transparency (EIT) spectra of Rydberg atoms. Using the Shirley method, we analyze the Floquet eigen-states to establish a direct connection between the measured spectra and the underlying dressed-state dynamics. By tracking the evolution of state mixing, bare-state composition, and dominant higher-order Floquet coupling pathways, we reveal the physical mechanisms responsible for the formation of avoided crossings across different principal quantum numbers and RF frequencies. Our results demonstrate that Floquet eigen-state analysis provides a powerful framework for interpreting complex AC Stark-shifted Rydberg-EIT spectra and the underlying higher-order Floquet coupling interactions.

\maketitle

%%%%%%%%%%%%%%%%%%%%% Introduction %%%%%%%%%%%%%%%%%%%%%%%%%%%%%%%

\section{Introduction}

The interaction of highly excited Rydberg atoms with radio-frequency (rf) and microwave electric fields has emerged as a powerful platform for precision electrometry \cite{kumar_rydberg-atom_2017, kitching_atom-based_2025, schlossberger_rydberg_2024}, quantum sensing \cite{norrgard_quantum_2021, schlossberger_primary_2025}, and the investigation of strongly driven quantum systems \cite{manchaiah_frequency_2026}.   Owing to their large principal quantum numbers ($n$), Rydberg states are exceptionally sensitive to external electric fields which enables the observation of rf-induced energy shifts, Autler–Townes splitting, and non-perturbative Stark effects over a broad frequency range \cite{song_continuous_2024}. Rydberg spectroscopy provides a direct, atom-based measurement of electric fields without the need for conventional calibration standards \cite{sedlacek_microwave_2012, holloway_atom-based_2017, holloway_broadband_2014, gallagher_rydberg_2005, tanasittikosol_microwave_2011}. 

%The resulting Rydberg-EIT technique has become an important tool for SI-traceable electric-field measurements and broadband RF sensing while simultaneously offering a convenient platform for studying the interaction of atoms with strong oscillating electromagnetic fields.

% Owing to their large principal quantum numbers (n), Rydberg states possess electric dipole moments that scale as $n^2$ and polarizabilities that approximately scale as $n^7$, making them exceptionally sensitive to external electric fields. 

%beyond the perturbative regime

\begin{figure}[!h]
    \centering
    \includegraphics[width=0.6\linewidth]{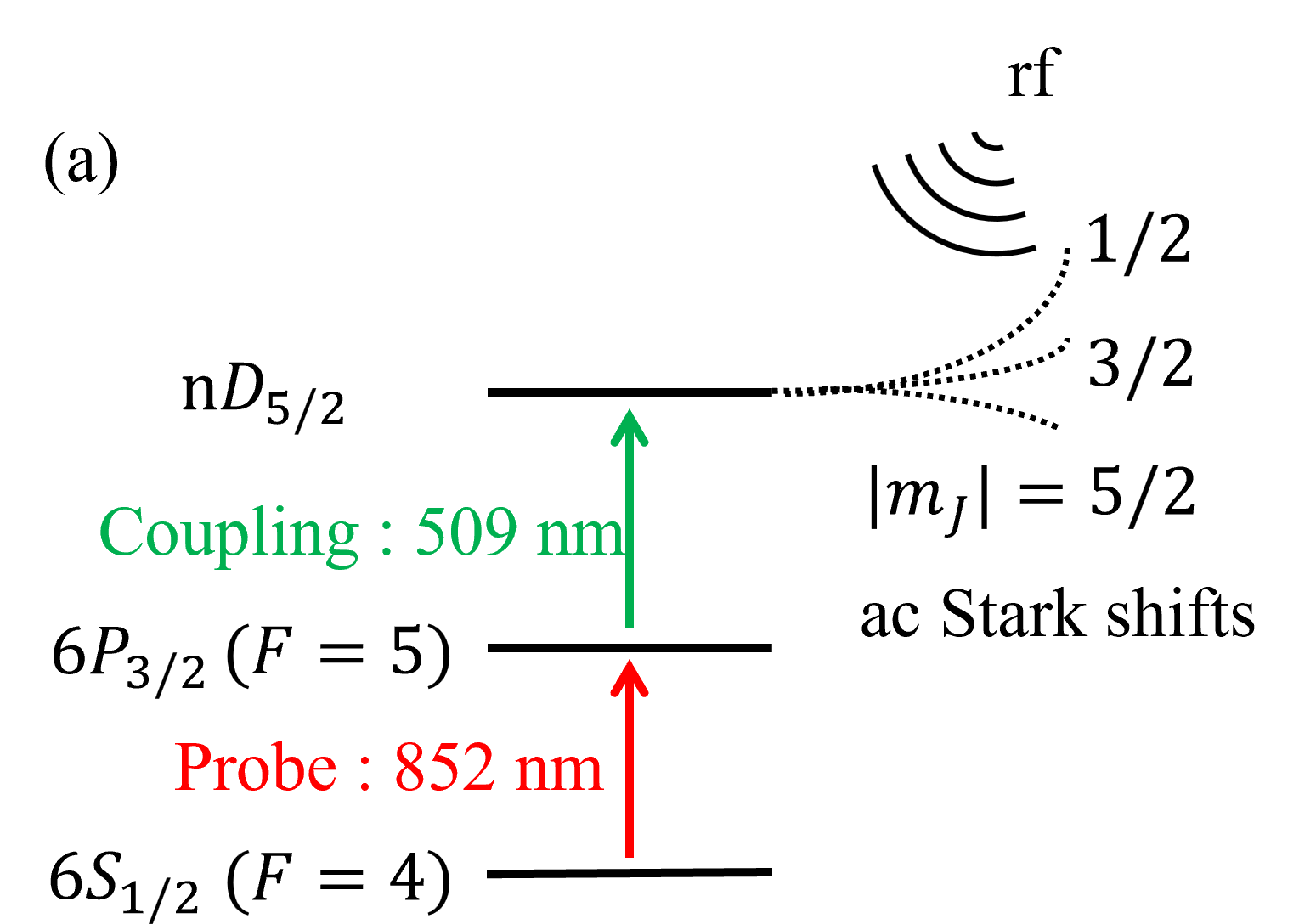}
    \includegraphics[width=\linewidth]{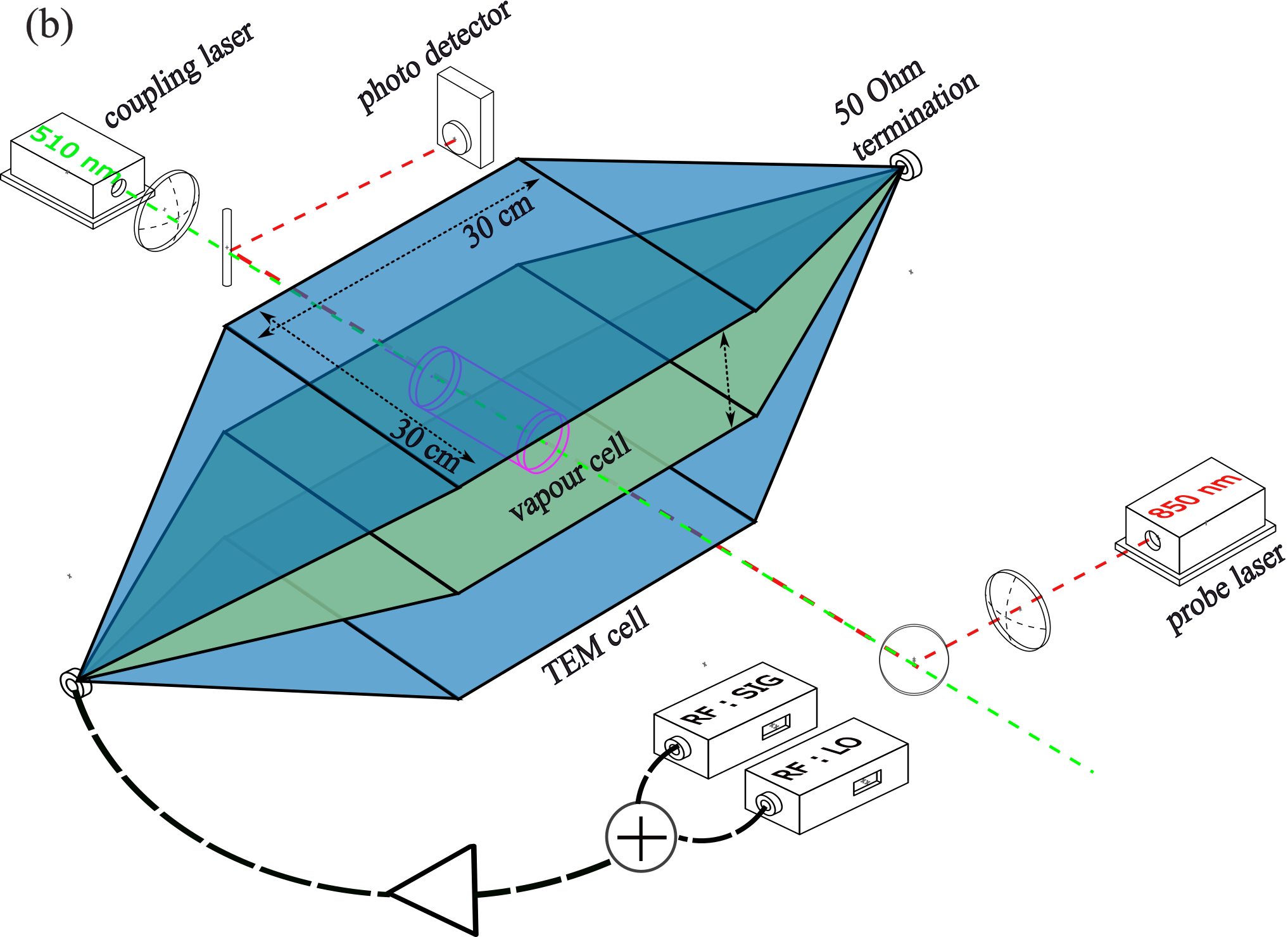}
    \caption{(a) Two-photon Rydberg excitation scheme with the Rydberg state undergoing Stark shifts under applied rf field and (b)  experimental setup used in this study. The cesium filled vapor cylindrical cell  is placed inside an open-wall transverse electromagnetic (TEM) cell driven by two rf sources for heterodyne detection. Their outputs are combined and amplified before being applied to the TEM cell. The coupling laser frequency is swept while the probe laser transmission through the vapor cell is detected with a photo-detector. By repeating the measurement for different applied rf electric field amplitudes, the AC Stark-shifted Rydberg EIT spectra are obtained.} 
    \label{fig:Levels_and_schematic}
\end{figure}

In the case of  non-resonant Rydberg atom sensing using Stark shifts \cite{chen_electric_2025, holloway_electromagnetically_2022, li_super_2023}, as the rf field strength increases, the interaction between neighboring Rydberg levels evolves  and gives rise to a rich Floquet energy spectrum characterized by numerous avoided crossings \cite{miller_radio-frequency-modulated_2016,paradis_atomic_2019,rotunno_detection_2023,jiao_atom-based_2017} which cannot be explained fully by conventional AC Stark maps alone. These avoided crossings occur when two Floquet states approach resonance and are coupled by a non-zero electric-dipole interaction, leading to level repulsion and coherent hybridization of the participating states. They provide direct insight into higher-order Floquet coupling interactions, rf polarization selection rules, and transition pathways.

Although the occurrence of avoided crossings is studied extensively \cite{wang_using_2016,feynman_quantum_2015,mcculloch_field_2017,wang_atom-interferometric_2015,oubre_avoided_2002,huang_robust_2018, ravets_coherent_2014,anderson_optical_2016, jiao_spectroscopy_2016,paradis_atomic_2019,noaman_rydberg-atom_2023,jiao_atom-based_2017}, their interpretation is often limited to identifying the associated spectral features in terms of even-order RF sidebands \cite{paradis_atomic_2019}, Floquet-based spectral calculations \cite{jiao_spectroscopy_2016, jiao_atom-based_2017,anderson_optical_2016}, and the Jaynes-Cummings framework \cite{noaman_rydberg-atom_2023}. These approaches successfully describe the positions and overall structure of the avoided crossings, but provide limited information about the bare state evolution and coupling pathways between the Floquet states. A more complete interpretation can be obtained by analyzing the evolution of the Floquet eigenstate instead of  considering the Stark spectrum alone. It provides the order of the coupling process, the coupling pathways through the intermediate Floquet states that lead to the  avoided crossing.

In this work, we present Rydberg-EIT measurements in the AC Stark-shift regime, together with Floquet-based Shirley-method \cite{shirley_solution_1965} calculations, to explain the avoided-crossing structures observed at different RF frequencies and electric-field strengths. A detailed analysis of the Floquet eigenstates in the vicinity of these avoided crossings provides insight into the RF-induced coupling between different Floquet states and the corresponding dominant coupling pathways. The evolution of state mixing across an avoided crossing is examined through the field-dependent contributions of the relevant bare Floquet basis states to the optically excited dressed state. This analysis reveals how the initially excited Rydberg state composition is redistributed among neighboring Floquet states as the RF field is increased. Through this, the intermediate states and the order of the coupling process are identified. In this paper, three representative cases exhibiting distinct coupling mechanisms and spectral characteristics are investigated, illustrating avoided crossings arising from different Floquet-state interactions and providing a systematic physical interpretation of their formation.

\begin{figure}
    \centering
    \includegraphics[width=\linewidth]{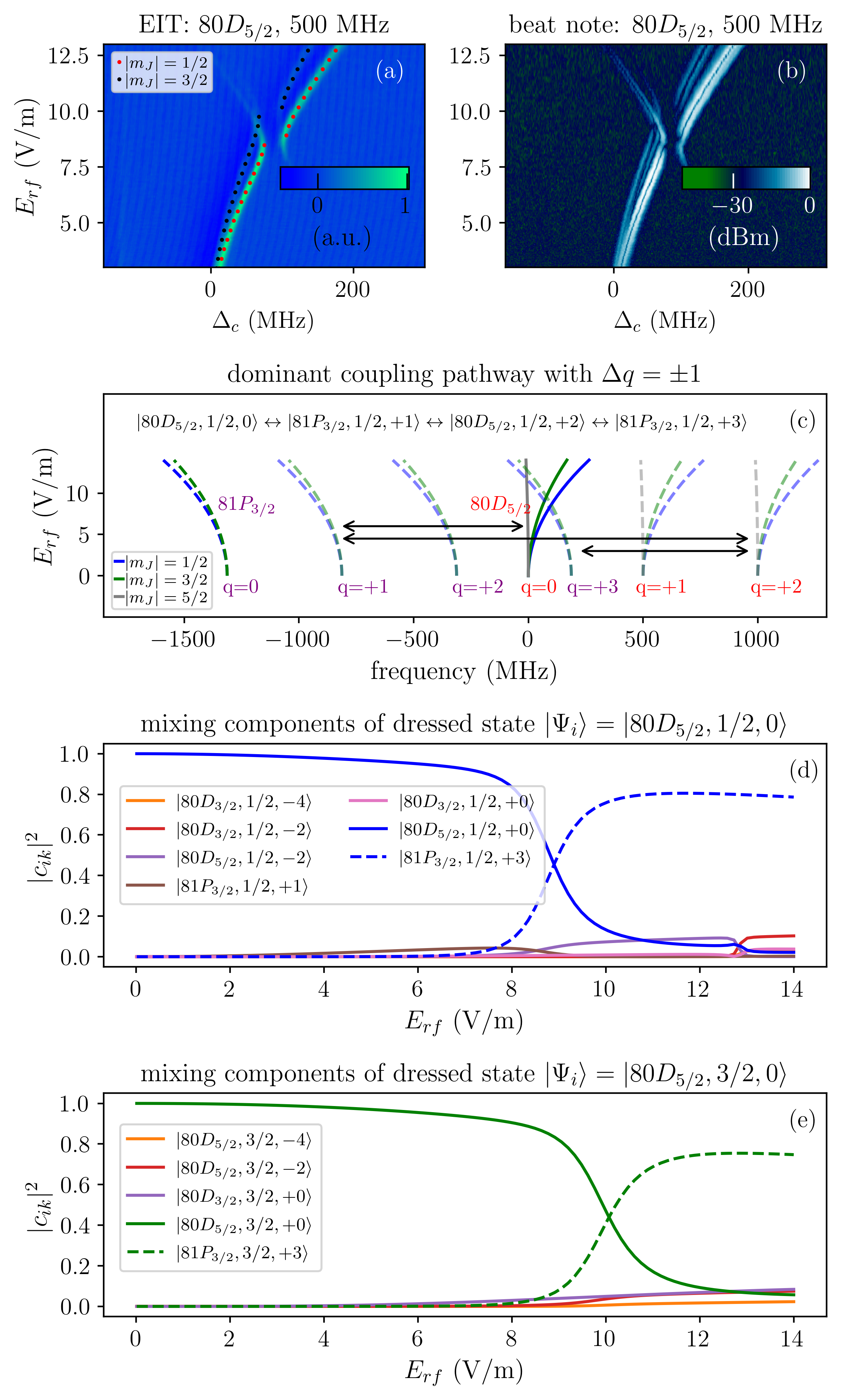}
    \caption{(a) Experimentally obtained AC Stark shifted EIT spectra of the $80D_{5/2}$ state as a function of the rf electric field magnitude at frequency of $500\,\mathrm{MHz}$ and the coupling detuning. (b) Concurrently obtained beat-note amplitude in the same 2D space.  (c) The higher-order Floquet coupling interpretation of the avoided crossings and sequential coupling of Floquet replicas. The lines with reduced opacity correspond to the Floquet replicas of the corresponding states. (d) Calculated contribution of the various bare states to the $m_J=1/2$ mixed state. The avoided crossing is the result of third order coupling between the states $\ket{80D_{5/2},1/2,+0}$ and $\ket{81P_{3/2},1/2,+3}$ at the applied field of approximately $8.8\,\mathrm{V/m}$.(e) The contribution of the various bare states to the $m_J=3/2$ mixed state. The avoided crossing is the result of third order coupling between the states $\ket{80D_{5/2},3/2,+0}$ and $\ket{81P_{3/2},3/2,+3}$ at the applied field of approximately $10\,\mathrm{V/m}$.}
    \label{fig:AC_n80_500MHz}
\end{figure}

\section{Experimental Setup}

In the two-photon Rydberg-excitation scheme shown in Fig.~\ref{fig:Levels_and_schematic}(a), the counter propagating probe and coupling lasers prepare a selected target Rydberg state in cesium atoms contained within a cylindrical vapor cell with 25 mm diameter  and 75 mm length  maintained at room temperature. Both laser beams have approximate  Full Width at Half Maximum (FWHM) diameter of $280\,\mathrm{\mu m}$, with optical powers of $56\,\mathrm{mW}$ and $18\,\mathrm{\mu W}$ for the coupling and probe beams, respectively. The lasers are linearly polarized and co-polarized with the RF fields. In the experimental setup shown in Fig.~\ref{fig:Levels_and_schematic}(b), the vapor cell is placed inside a TEM cell driven by two rf sources, signal (sig) and Local Oscillator (LO), through an amplifier. The two rf tones are separated by $200\,\mathrm{kHz}$, which defines the beat-note frequency detected in the transmitted probe signal. This approach \cite{jing_atomic_2020, simons_rydberg_2019} enables the extraction of the amplitude and phase information of an rf signal using Rydberg-atom spectroscopy. Typically, the signal rf source  is maintained at a fixed low power, whereas the power of the LO, offset in frequency by the beat-note frequency, is swept to determine the optimum operating point in the two-dimensional parameter space of coupling-laser detuning $(\Delta_c)$ and rf power. The frequency dependent calibration factor relating the source power to the effective electric field amplitude experienced by the atoms is determined using AC Stark shifting method. The low power rf source corresponds to an effective field of $63\,\mathrm{mV/m}$ at the atoms. The electric field amplitude range associated with the swept RF source is indicated in the avoided crossings plots.

The target Rydberg state becomes dressed by the rf interaction and optical transition strength to a given dressed state is approximately proportional to its overlap with the optically accessible bare Rydberg state. As the rf-field amplitude increases, the overlap of a particular dressed state with the original target state may decrease, while other dressed states acquire greater overlap. Consequently, the EIT spectral weight is redistributed among multiple dressed-state resonances, resulting in changes in the peak amplitudes and the appearance of additional spectral features.

\section{Shirley Method}
%The interaction of a Rydberg atom with a monochromatic rf electric field is described the time-dependent Hamiltonian \begin{equation}
%    H(t) = H_0 - \mathbf{d} \cdot \mathbf{E}_{rf} \cos(\omega t)
%    \label{eq:Time-Hamiltonian}
%\end{equation}
%where $H_0$ is the field-free atomic Hamiltonian, $\mathbf{d}$ is the electric dipole operator,  and $\mathbf{E}_{rf}$ is the rf electric field amplitude. 

%The method demonstrates that the solution of the time-dependent Schrödinger equation with a periodic Hamiltonian is equivalent to the solution of another Schrödinger equation with a time-independent Hamiltonian represented by an infinite-dimensional matrix. 

The Shirley method \cite{shirley_solution_1965} is one of the procedures available in  the alkali Rydberg Calculator (ARC) python package \cite{sibalic_arc_2017} for obtaining Stark maps of Rydberg states driven by an alternating-current (AC) electric field. The method demonstrates that the solution of the time-dependent Schrödinger equation with a periodic Hamiltonian can be transformed into an equivalent problem involving a time-independent Hamiltonian represented by an infinite-dimensional matrix. This equivalence transforms the problem into an eigenvalue problem, where the eigenvalues and eigenvectors of the infinite-dimensional Hamiltonian are determined, from which the evolution operator can be expressed. The equivalent time-independent Shirley Hamiltonian is
\begin{equation}
    H = H_0 + f dT + E_{rf} B,
    \label{eq:Shirley_Hamiltonian}
\end{equation}
where $H_0$ contains the field-free atomic energies. The diagonal matrix $dT$ represents the Floquet frequency offsets, and the term $f dT$ generates replicated Floquet manifolds separated according to the applied RF frequency $f$. The matrix $B$ is the RF coupling matrix that describes the electric-dipole coupling between these Floquet manifolds, leading to RF-induced state mixing. The coupling strength is scaled by the applied rf electric-field amplitude $E_{rf}$. This notation follows the implementation of the Shirley method in the ARC package\cite{sibalic_arc_2017}.

\begin{figure}
    \centering
    \includegraphics[width=\linewidth]{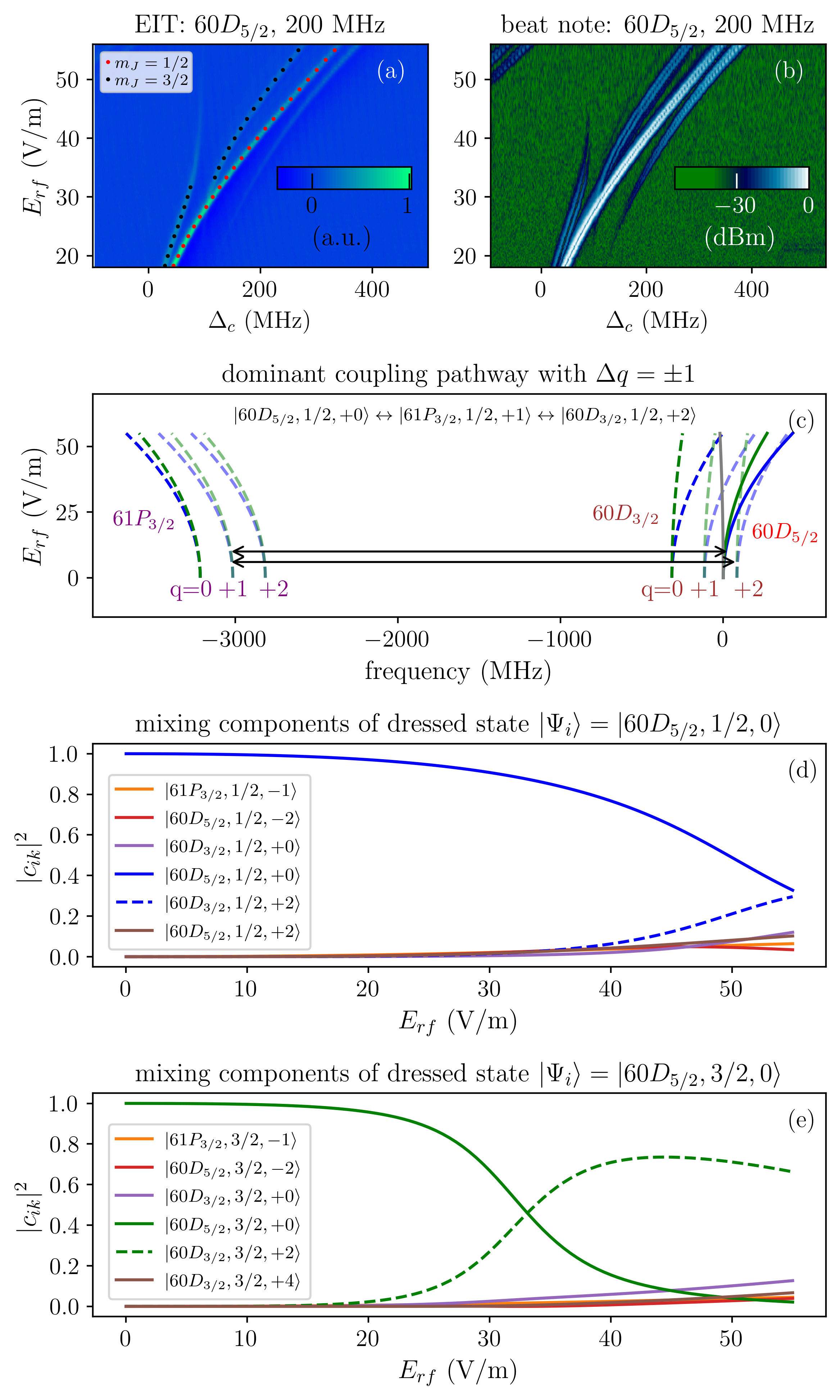}
      \caption{(a) AC Stark shifted EIT spectra of the $60D_{5/2}$ at frequency of $200\,\mathrm{MHz}$ (b) Beat-note amplitude  (c) Coupling pathway interpretation  (d)  $m_J=1/2$ mixed state composition. (e) $m_J=3/2$ mixed state composition. Avoided crossing due to second order coupling between the states $\ket{60D_{5/2},3/2,+0}$ and  $\ket{60D_{3/2},3/2,+2}$ is at approximate field strength of $33\,\mathrm{V/m}$}.
    \label{fig:AC_n60_200MHz}
\end{figure}

%where $H_0$ is the bare Hamiltonian that determines the field-free atomic energies.  $dT$ is the Floquet frequency matrix which contains the Floquet energy offsets that generate the replicated Floquet manifolds  corresponding to the rf frequency $f$. $B$ is the rf coupling matrix that couples the replicated manifolds through electric-dipole interactions that produce the mixing of rf-induced state. It is scaled by the applied rf electric field amplitude $E_{rf}$.

 Let the manifold of bare Rydberg states of interest be represented by $\ket{nl_j,m_J}$. The Shirley method defines the bare Floquet states $\ket{nl_j,m_J,q}$ where $q$ indexes the Fourier replica of the periodically driven  quasi-energy state.In the convention used here, $q=+1$ represents the upward energy shift by one rf-frequency interval, whereas $q=-1$ represents the corresponding downward shift; similarly, $q=\pm n$ denotes shifts by $\pm n$ rf-frequency intervals. It provides the output of the periodic Schrödinger equation in terms of eigenvalues, eigenvectors, target shifts, and transition probabilities. Target shifts are the shifts of the states relative to the zero field energy for an applied field. In contrast, eigenvalues correspond to the energies of the dressed Floquet states. Target shifts are more directly comparable with experimental observations because they represent the evolution of the experimentally addressed bare state. %Although the Shirley method also provides transition probabilities, they are not considered in this work because they quantify the coupling strengths between directly connected atomic basis states rather than between dressed Floquet states.

For interpreting the results of avoided crossings, understanding the composition of the mixed states is essential. The mixed Floquet states are connected through a chain of  couplings involving intermediate states, with each coupling represented by the interaction term $E_{rf}B$ in the Shirley Hamiltonian of Eq.\,\ref{eq:Shirley_Hamiltonian}. Analyzing the evolution of the eigenvector weights as a function of the rf electric-field  amplitude provides a direct means of identifying the bare Floquet states that participate in the coupling.  It also explains why certain resonances exhibit pronounced avoided crossings, whereas others remain nearly uncoupled because the effective coupling between the adjacent Floquet states is either too weak or forbidden by the applicable selection rules. Suppose that a Floquet eigen-state $\ket{\Psi_i}$ is expanded as
\begin{equation}
\ket{\Psi_i} = \sum_k c_{ik} \ket{k},
\end{equation}
where $\ket{k}$ denotes a bare Floquet basis state $\ket{n,l,j,m_J,q}$, and the coefficient $c_{ik}$ specifies the amplitude of that basis state in the eigen-state $\ket{\Psi_i}$.  For a normalized eigen-state, $|c_{ik}|^2$ represents the corresponding basis-state weight, or equivalently the probability of obtaining the state $\ket{k}$ upon projection onto the bare Floquet basis. The largest values of $|c_{ik}|^2$ indicate which bare Floquet basis states contribute most strongly to the dressed state. As the rf electric-field amplitude increases, the eigen-states may undergo substantial mixing, indicating that a given dressed state can no longer be identified with a single bare atomic level.

%The Floquet eigen states are the physical dressed atom-field states that correspond, in the classical field limit, to the dressed states obtained from a fully quantized atom-plus-field Hamiltonian. The Floquet/ Shirley method captures the effects of repeated absorption and stimulated emission of rf quanta through the coupling between different $q$ sectors, without ever explicitly introducing photon-number states.

The eigenvectors obtained from the Shirley diagonalization become a mixture of several Floquet basis states, as suggested by Eq.~\ref{eq:Shirley_Hamiltonian}. The EIT peak amplitude associated with each branch depends on the contribution of the optically excited bare state to that mixed state. Near an avoided crossing, the EIT intensity is transferred from one branch to another, while the corresponding energy branches repel instead of crossing. This transfer of EIT intensity between the two branches indicates the mixing of the participating states. The evolution of this state mixing across the avoided crossing is analyzed in the next section in three different cases. In the Shirley-method calculations \cite{sibalic_arc_2017} presented in this paper, the basis includes states within ($\pm 5$) of the target principal quantum number, with ($l<5$) and Floquet indices up to ($q=\pm 5$). Convergence was verified by varying these basis limits and confirming that the calculated results remained unchanged.

\section{Results and Discussion}

% =======================================================================================

\begin{figure}
    \centering
    \includegraphics[width=\linewidth]{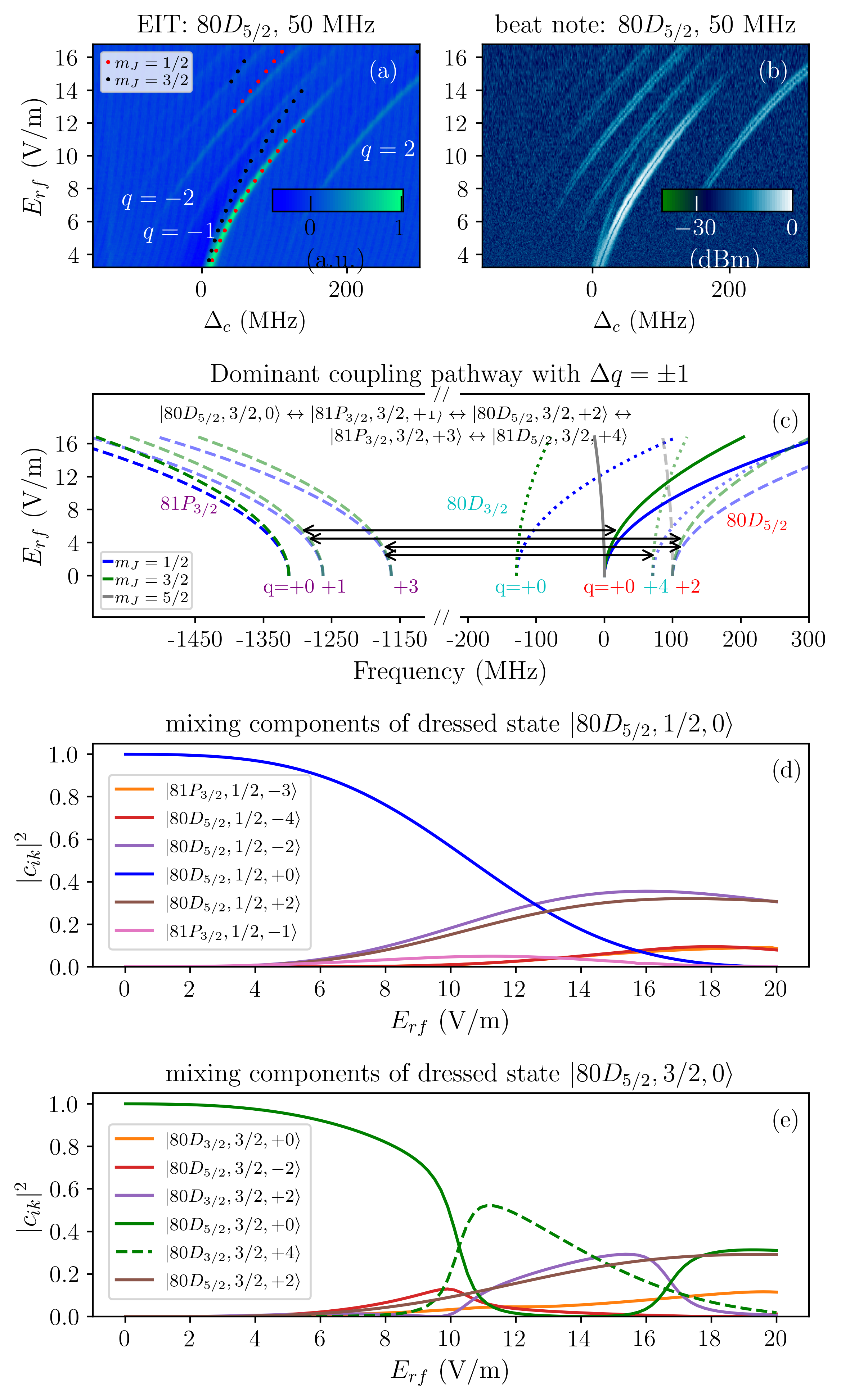}
    \caption{(a) AC Stark shifted EIT spectra of the $80D_{5/2}$ at frequency of $50\,\mathrm{MHz}$ (b) Beat-note amplitude  (c) Coupling pathway interpretation. (d)  $m_J=1/2$ mixed state composition. The Floquet replicas of the state $\ket{80D_{5/2},m_J=1/2}$ at $q=-4,-2,0,+2$ show up in the calculations. However, the $q=-1,-3$ replicas visible in the experimental results are not indicated in the Shirley calculations. (e) $m_J=3/2$ mixed state composition. Avoided crossing at $10\,\mathrm{V/m}$ can be observed in this case.}
    \label{fig:AC_n80_50MHz}
\end{figure}

To demonstrate the evolution of state mixing and the resulting avoided crossings, three examples exhibiting distinct characteristics in their EIT spectra are presented in this work. Figure~\ref{fig:AC_n80_500MHz} shows the results for excitation to the $80D_{5/2}$ Rydberg state under an applied rf LO field with a frequency of $500\,\mathrm{MHz}$. In Fig.~\ref{fig:AC_n80_500MHz}(a), the experimentally measured AC Stark-shifted EIT spectrum, obtained by varying the amplitude of the applied rf electric field, is presented. The experimental spectrum is overlaid with Shirley-method calculations of the energy shifts associated with the $m_J=1/2$ and $m_J=3/2$ sublevels showing perfect agreement. Because the linearly polarized probe and coupling laser fields are aligned with the rf field inside the vapor cell, the $m_J=5/2$ sublevel is only weakly populated and is therefore not resolved in the measured spectrum.

%The beat note generated confirms the character of the dressed state jumping across the avoided crossings.
In Fig.~\ref{fig:AC_n80_500MHz}(b), the corresponding beat-note spectrum is shown in the same two-dimensional parameter space defined by the applied rf electric-field amplitude and the coupling-laser detuning. The redistribution of the beat-note amplitude across the avoided crossing confirms the transfer of the dressed-state character from one branch to the other. In Fig.~\ref{fig:AC_n80_500MHz}(c), the calculated AC Stark-shifted energies of the $80D_{5/2}$ and $81P_{3/2}$ states, together with their relevant Floquet replicas with $q=1,2,3$, are plotted. The sequential coupling pathway responsible for the observed avoided crossing between $\ket{80D_{5/2},m_J=1/2,q=0}$ and $\ket{81P_{3/2},m_J=1/2,q=+3}$ is also indicated. It preserves the parity change ($\Delta l = \pm 1$) associated with each single electric dipole transition. An analogous interpretation applies to the $m_J=3/2$ manifold.
This example demonstrates an avoided crossing resulting from the interaction between a $D$ state with  $n=80$ and a Floquet replica associated with a $P$ state having $n=81$.
%This example shows the avoided crossing resulting from the interaction of D state of one n=80 with the Floquet state corresponding to P state of another n=81.

In Fig.~\ref{fig:AC_n80_500MHz}(d), the Shirley-method analysis of state mixing is presented through the fractional contributions of the relevant Floquet basis states to the dressed state that evolves from $\ket{80D_{5/2},m_J=1/2,q=0}$ as the rf electric-field amplitude is increased. The contributions are normalized over the complete basis and only the Floquet states with appreciable contributions over the investigated field range are included. At low rf electric-field amplitudes, the dressed state remains nearly pure and is dominated by the $\ket{80D_{5/2},m_J=1/2,q=0}$ basis component. However, at field amplitudes above approximately $8\,\mathrm{V/m}$, the contribution from $\ket{81P_{3/2},m_J=1/2,q=+3}$ increases substantially, producing strong state mixing and the associated avoided crossing. The much smaller contributions from the remaining Floquet states are neglected in the present interpretation. Similarly, Fig.~\ref{fig:AC_n80_500MHz}(e) shows the corresponding state-mixing behavior for the $m_J=3/2$ manifold.

Figure~\ref{fig:AC_n60_200MHz} shows an avoided crossing involving the $60D_{5/2}$ state at an rf frequency of $200\,\mathrm{MHz}$. Unlike the example in Fig.~\ref{fig:AC_n80_500MHz}, this avoided crossing results from coupling to a Floquet replica of another $D$ state within the same principal-quantum-number manifold, $n=60$.  The remaining features of the figure can be interpreted in the same manner as those described for Fig.~\ref{fig:AC_n80_500MHz}. This case is particularly important because a direct electric dipole transition between the two $D$ states is forbidden. The observed avoided crossing therefore arises through intermediate coupling with a $P$ state. This also explains why a similar avoided crossing does not occur at $400\,\mathrm{MHz}$, even though twice the RF frequency matches the energy separation between the two $D$ states, because the required direct $D\leftrightarrow D$ is not allowed. This confirms that the appropriate energy separation alone does not guarantee the occurrence of an avoided crossing.

%This microscopic analysis explains and helps in understanding why the avoided crossings occur at specific frequencies and states with specific principal quantum numbers. It helps in understanding the complex spectroscopic features which emerge in the EIT Rydberg sensing applications.

This microscopic analysis provides a physical explanation for why avoided crossings appear only at particular RF frequencies and for specific combinations of Rydberg states. The complex spectroscopic features encountered in Rydberg-EIT experiments can be interpreted from participating Floquet states and the associated coupling pathways.

%In Fig.\ref{fig:AC_n60_200MHz}, the avoided crossing of the state $60D_{5/2}$ at rf frequency of $200\,\mathrm{MHz}$ resulting from a Floquet replica of a $D$ state of same $n$ is presented. Same decription of Fig.\ref{fig:AC_n80_500MHz} applies in this case too.

% =====================================================================================

%The results shown in Fig.~\ref{fig:AC_n80_50MHz} represent a qualitatively different case from those described above, in which multiple Floquet replicas are simultaneously resolved in the EIT spectrum. It represents the case of $80D_{5/2}$ state driven with the RF frequency of $50\,\mathrm{MHz}.$ For the $m_J=1/2$ manifold, contributions from the Floquet replicas with $q=-4,-3,-2,-1,$ and $+2$ become evident at higher rf electric-field amplitudes. The contributions of these Floquet replica states to the dressed state can be seen in Fig.\ref{fig:AC_n80_50MHz}(d). However, the Shirley method calcualtions do not reflect the $q=-1$ and $q=-3$ replicas. These could be due to a stray electric field \cite{rotunno_detection_2023}. By contrast, the $m_J=3/2$ state exhibits a weak avoided crossing induced by the $q=+4$ Floquet replica of the nearby $80D_{3/2}$ state. 

The results shown in Fig.~\ref{fig:AC_n80_50MHz} represent a qualitatively different case from those described above, in which multiple Floquet replicas are simultaneously resolved in the EIT spectrum. This case corresponds to the $80D_{5/2}$ state driven by an RF field at $50\,\mathrm{MHz}.$ For the $m_J=1/2$ manifold, contributions from the Floquet replicas with $q=-4,-3,-2,-1,$ and $+2$ become evident at higher rf electric-field amplitudes. The contributions of only some of these Floquet basis  states to the dressed state can be seen in Fig.\ref{fig:AC_n80_50MHz}(d). Shirley-method calculations do not reproduce the experimentally observed $q=-1$ and $q=-3$ replicas. These additional features may arise from stray electric fields or other symmetry-breaking perturbations not included in the present model \cite{rotunno_detection_2023}. By contrast, the $m_J=3/2$ state exhibits a weak avoided crossing induced by the $q=+4$ Floquet replica of the nearby $80D_{3/2}$ state as evident in Fig.\,\ref{fig:AC_n80_50MHz}(e). The dominant coupling pathway associated with this avoided crossing is
\begin{align*}
\ket{80D_{5/2},3/2,+0}&\leftrightarrow\ket{81P_{3/2},3/2,+1}\leftrightarrow\\\ket{80D_{5/2},3/2,+2}&\leftrightarrow\ket{81P_{3/2},3/2,+3}\leftrightarrow\ket{80D_{3/2},3/2,+4}.
\end{align*}

Many alternative  coupling pathways are also possible, including for cases discussed previously. However, such pathways  contribute less strongly because the product of the dipole matrix elements along the successive coupling steps is smaller and the detuning along the pathways is higher. %A quantitative ranking of all possible pathways according to their effective end-to-end coupling strengths is beyond the scope of the present work.

One direct consequence of encountering avoided crossings is that, if the coupling laser is locked near an avoided crossing region, the beat-note signal may disappear over certain ranges of the applied LO field amplitude. For the case shown in Fig.\ref{fig:AC_n80_500MHz}, when the coupling laser is locked at a detuning of approximately $70\,\mathrm{MHz}$, the EIT response does not provide a suitable operating point over the corresponding LO-field range, and consequently no detectable beat-note signal is obtained.

\section{Conclusion}

The experimentally observed avoided crossings in Rydberg-EIT spectra at selected rf frequencies and field amplitudes are interpreted using the Shirley method as coupling between Rydberg states and their Floquet replicas. Comparison of the measured spectra with calculated Floquet energies and eigenvectors identifies the participating states, reveals the evolution of state mixing, and provides insights into the coupling of adjacent Floquet states. The analysis also explains why some resonances exhibit pronounced avoided crossings, whereas others remain weakly coupled because of small effective coupling strengths or dipole-selection rules. These results demonstrate that the Shirley method provides not only AC Stark maps but also a detailed physical interpretation of complex rf-dressed Rydberg-EIT spectra. 

Applications  of avoided crossings include strong-field measurements \cite{anderson_optical_2016} and their use as RF electric field calibration markers \cite{jiao_atom-based_2017}. Future studies could explore alternative laser and rf polarizations, higher-angular-momentum states $(l>2)$, and magnetic-field-induced modifications of the avoided crossings for potential magnetometry applications.

\section*{data availability}
   All of the data presented in this paper and used to
support the conclusions of this article is available at \cite{prajapati_data_nodate}.

\begin{acknowledgments}
A contribution of the U.S. government, this work is not subject to copyright in the U.S.
\end{acknowledgments}
\bibliography{Refs}% Produces the bibliography via BibTeX.

@article{song_continuous_2024,
    title = {Continuous broadband {Rydberg} receiver using {AC} {Stark} shifts and {Floquet} states},
    volume = {125},
    issn = {0003-6951},
    url = {https://doi.org/10.1063/5.0227250},
    doi = {10.1063/5.0227250},
    number = {19},
    urldate = {2026-06-20},
    journal = {Applied Physics Letters},
    author = {Song, Danni and Jiao, Yuechun and Hu, Jinlian and Yin, Yuwen and Li, Zhenhua and He, Yunhui and Bai, Jingxu and Zhao, Jianming and Jia, Suotang},
    month = nov,
    year = {2024},
    pages = {194001},
}

@article{jiao_atom-based_2017,
    title = {Atom-{Based} {Radio}-{Frequency} {Field} {Calibration} and {Polarization} {Measurement} {Using} {Cesium} n {D} {J} {Floquet} {States}},
    volume = {8},
    copyright = {http://link.aps.org/licenses/aps-default-license},
    issn = {2331-7019},
    url = {http://link.aps.org/doi/10.1103/PhysRevApplied.8.014028},
    doi = {10.1103/PhysRevApplied.8.014028},
    number = {1},
    urldate = {2026-05-05},
    journal = {Physical Review Applied},
    author = {Jiao, Yuechun and Hao, Liping and Han, Xiaoxuan and Bai, Suying and Raithel, Georg and Zhao, Jianming and Jia, Suotang},
    month = jul,
    year = {2017},
    pages = {014028},
}

@article{miller_radio-frequency-modulated_2016,
    title = {Radio-frequency-modulated {Rydberg} states in a vapor cell},
    volume = {18},
    issn = {1367-2630},
    url = {https://doi.org/10.1088/1367-2630/18/5/053017},
    doi = {10.1088/1367-2630/18/5/053017},
    number = {5},
    urldate = {2026-02-03},
    journal = {New Journal of Physics},
    publisher = {IOP Publishing},
    author = {Miller, S A and Anderson, D A and Raithel, G},
    month = may,
    year = {2016},
    pages = {053017},
}

@article{paradis_atomic_2019,
    title = {Atomic measurements of high-intensity {VHF}-band radio-frequency fields with a {Rydberg} vapor-cell detector},
    volume = {100},
    url = {https://link.aps.org/doi/10.1103/PhysRevA.100.013420},
    doi = {10.1103/PhysRevA.100.013420},
    number = {1},
    urldate = {2026-02-02},
    journal = {Physical Review A},
    publisher = {American Physical Society},
    author = {Paradis, Eric and Raithel, Georg and Anderson, David A.},
    month = jul,
    year = {2019},
    pages = {013420},
}

@article{rotunno_detection_2023,
    title = {Detection of 3–300 {MHz} electric fields using {Floquet} sideband gaps by “{Rabi} matching” dressed {Rydberg} atoms},
    volume = {134},
    issn = {0021-8979, 1089-7550},
    url = {https://pubs.aip.org/jap/article/134/13/134501/2914063/Detection-of-3-300-MHz-electric-fields-using},
    doi = {10.1063/5.0162101},
    number = {13},
    urldate = {2024-09-11},
    journal = {Journal of Applied Physics},
    author = {Rotunno, Andrew P. and Berweger, Samuel and Prajapati, Nikunjkumar and Simons, Matthew T. and Artusio-Glimpse, Alexandra B. and Holloway, Christopher L. and Jayaseelan, Maitreyi and Potvliege, R. M. and Adams, C. S.},
    month = oct,
    year = {2023},
    pages = {134501},
}

@article{shirley_solution_1965,
    title = {Solution of the {Schrödinger} {Equation} with a {Hamiltonian} {Periodic} in {Time}},
    volume = {138},
    copyright = {http://link.aps.org/licenses/aps-default-license},
    issn = {0031-899X},
    url = {https://link.aps.org/doi/10.1103/PhysRev.138.B979},
    doi = {10.1103/PhysRev.138.B979},
    number = {4B},
    urldate = {2026-03-30},
    journal = {Physical Review},
    author = {Shirley, Jon H.},
    month = may,
    year = {1965},
    pages = {B979--B987},
}

@article{sedlacek_microwave_2012,
    title = {Microwave electrometry with {Rydberg} atoms in a vapour cell using bright atomic resonances},
    volume = {8},
    copyright = {2012 Springer Nature Limited},
    issn = {1745-2481},
    url = {https://www.nature.com/articles/nphys2423},
    doi = {10.1038/nphys2423},
    number = {11},
    urldate = {2024-09-11},
    journal = {Nature Physics},
    publisher = {Nature Publishing Group},
    author = {Sedlacek, Jonathon A. and Schwettmann, Arne and Kübler, Harald and Löw, Robert and Pfau, Tilman and Shaffer, James P.},
    month = nov,
    year = {2012},
    pages = {819--824},
}

@article{holloway_atom-based_2017,
    title = {Atom-{Based} {RF} {Electric} {Field} {Metrology}: {From} {Self}-{Calibrated} {Measurements} to {Subwavelength} and {Near}-{Field} {Imaging}},
    volume = {59},
    issn = {1558-187X},
    shorttitle = {Atom-{Based} {RF} {Electric} {Field} {Metrology}},
    url = {https://ieeexplore.ieee.org/abstract/document/7812705},
    doi = {10.1109/TEMC.2016.2644616},
    number = {2},
    urldate = {2024-09-10},
    journal = {IEEE Transactions on Electromagnetic Compatibility},
    author = {Holloway, Christopher L. and Simons, Matthew T. and Gordon, Joshua A. and Wilson, Perry F. and Cooke, Caitlyn M. and Anderson, David A. and Raithel, Georg},
    month = apr,
    year = {2017},
    note = {Conference Name: IEEE Transactions on Electromagnetic Compatibility},
    pages = {717--728},
}

@article{holloway_broadband_2014,
    title = {Broadband {Rydberg} {Atom}-{Based} {Electric}-{Field} {Probe} for {SI}-{Traceable}, {Self}-{Calibrated} {Measurements}},
    volume = {62},
    issn = {1558-2221},
    url = {https://ieeexplore.ieee.org/document/6910267},
    doi = {10.1109/TAP.2014.2360208},
    number = {12},
    urldate = {2024-09-11},
    journal = {IEEE Transactions on Antennas and Propagation},
    author = {Holloway, Christopher L. and Gordon, Joshua A. and Jefferts, Steven and Schwarzkopf, Andrew and Anderson, David A. and Miller, Stephanie A. and Thaicharoen, Nithiwadee and Raithel, Georg},
    month = dec,
    year = {2014},
    note = {Conference Name: IEEE Transactions on Antennas and Propagation},
    pages = {6169--6182},
}

@article{sibalic_arc_2017,
    title = {{ARC}: {An} open-source library for calculating properties of alkali {Rydberg} atoms},
    volume = {220},
    issn = {0010-4655},
    shorttitle = {{ARC}},
    url = {https://www.sciencedirect.com/science/article/pii/S0010465517301972},
    doi = {10.1016/j.cpc.2017.06.015},
    urldate = {2025-11-13},
    journal = {Computer Physics Communications},
    author = {Šibalić, N. and Pritchard, J. D. and Adams, C. S. and Weatherill, K. J.},
    month = nov,
    year = {2017},
    pages = {319--331},
}

@book{gallagher_rydberg_2005,
    address = {Cambridge},
    edition = {Digitally print. 1. pbk version},
    series = {Cambridge monographs on atomic, molecular, and chemical physics},
    title = {Rydberg atoms},
    isbn = {978-0-521-02166-1 978-0-521-38531-2},
    number = {3},
    publisher = {Cambridge Univ. Press},
    author = {Gallagher, Thomas F.},
    year = {2005},
}

@article{tanasittikosol_microwave_2011,
    title = {Microwave dressing of {Rydberg} dark states},
    volume = {44},
    issn = {0953-4075},
    url = {https://doi.org/10.1088/0953-4075/44/18/184020},
    doi = {10.1088/0953-4075/44/18/184020},
    number = {18},
    urldate = {2026-07-22},
    journal = {Journal of Physics B: Atomic, Molecular and Optical Physics},
    author = {Tanasittikosol, M and Pritchard, J D and Maxwell, D and Gauguet, A and Weatherill, K J and Potvliege, R M and Adams, C S},
    month = sep,
    year = {2011},
    pages = {184020},
}

@article{kumar_rydberg-atom_2017,
    title = {Rydberg-atom based radio-frequency electrometry using frequency modulation spectroscopy in room temperature vapor cells},
    volume = {25},
    copyright = {© 2017 Optical Society of America},
    issn = {1094-4087},
    url = {https://opg.optica.org/oe/abstract.cfm?uri=oe-25-8-8625},
    doi = {10.1364/OE.25.008625},
    number = {8},
    urldate = {2025-08-06},
    journal = {Optics Express},
    publisher = {Optica Publishing Group},
    author = {Kumar, Santosh and Fan, Haoquan and Kübler, Harald and Jahangiri, Akbar J. and Shaffer, James P.},
    month = apr,
    year = {2017},
    pages = {8625--8637},
}

@article{jing_atomic_2020,
    title = {Atomic superheterodyne receiver based on microwave-dressed {Rydberg} spectroscopy},
    volume = {16},
    copyright = {2020 The Author(s), under exclusive licence to Springer Nature Limited},
    issn = {1745-2481},
    url = {https://www.nature.com/articles/s41567-020-0918-5},
    doi = {10.1038/s41567-020-0918-5},
    number = {9},
    urldate = {2024-09-10},
    journal = {Nature Physics},
    publisher = {Nature Publishing Group},
    author = {Jing, Mingyong and Hu, Ying and Ma, Jie and Zhang, Hao and Zhang, Linjie and Xiao, Liantuan and Jia, Suotang},
    month = sep,
    year = {2020},
    pages = {911--915},
}

@article{simons_rydberg_2019,
    title = {A {Rydberg} atom-based mixer: {Measuring} the phase of a radio frequency wave},
    volume = {114},
    issn = {0003-6951, 1077-3118},
    shorttitle = {A {Rydberg} atom-based mixer},
    url = {https://pubs.aip.org/apl/article/114/11/114101/36735/A-Rydberg-atom-based-mixer-Measuring-the-phase-of},
    doi = {10.1063/1.5088821},
    number = {11},
    urldate = {2024-08-13},
    journal = {Applied Physics Letters},
    author = {Simons, Matthew T. and Haddab, Abdulaziz H. and Gordon, Joshua A. and Holloway, Christopher L.},
    month = mar,
    year = {2019},
    pages = {114101},
}

@article{anderson_optical_2016,
    title = {Optical {Measurements} of {Strong} {Microwave} {Fields} with {Rydberg} {Atoms} in a {Vapor} {Cell}},
    volume = {5},
    url = {https://link.aps.org/doi/10.1103/PhysRevApplied.5.034003},
    doi = {10.1103/PhysRevApplied.5.034003},
    number = {3},
    urldate = {2026-08-17},
    journal = {Physical Review Applied},
    publisher = {American Physical Society},
    author = {Anderson, D. A. and Miller, S. A. and Raithel, G. and Gordon, J. A. and Butler, M. L. and Holloway, C. L.},
    month = mar,
    year = {2016},
    pages = {034003},
}

@article{wang_using_2016,
    title = {Using experimental measurements of three-level avoided crossings to determine the quantum defect for the {Stark} map of highly excited cesium {Rydberg} atoms},
    volume = {93},
    url = {https://link.aps.org/doi/10.1103/PhysRevA.93.033416},
    doi = {10.1103/PhysRevA.93.033416},
    number = {3},
    urldate = {2026-08-17},
    journal = {Physical Review A},
    publisher = {American Physical Society},
    author = {Wang, Limei and Li, Changyong and Zhang, Hao and Zhang, Linjie and Yang, Yonggang and Man, Yuan and Zhao, Jianming and Jia, Suotang},
    month = mar,
    year = {2016},
    pages = {033416},
}

@article{oubre_avoided_2002,
    title = {Avoided {Crossings} in the {Interaction} of a {Xe} {Rydberg} {Atom} with a {Metal} {Surface}},
    volume = {106},
    issn = {1520-6106},
    url = {https://doi.org/10.1021/jp025875+},
    doi = {10.1021/jp025875+},
    number = {33},
    urldate = {2026-08-17},
    journal = {The Journal of Physical Chemistry B},
    author = {Oubre, C. and Nordlander, P. and Dunning, F. B.},
    month = jun,
    year = {2002},
    pages = {8338--8341},
}

@article{huang_robust_2018,
    title = {Robust {Rydberg} gate via {Landau}-{Zener} control of {F}{\textbackslash}"orster resonance},
    volume = {98},
    url = {https://link.aps.org/doi/10.1103/PhysRevA.98.052324},
    doi = {10.1103/PhysRevA.98.052324},
    number = {5},
    urldate = {2026-08-17},
    journal = {Physical Review A},
    publisher = {American Physical Society},
    author = {Huang, Xi-Rong and Ding, Zong-Xing and Hu, Chang-Sheng and Shen, Li-Tuo and Li, Weibin and Wu, Huaizhi and Zheng, Shi-Biao},
    month = nov,
    year = {2018},
    pages = {052324},
}

@article{feynman_quantum_2015,
    title = {Quantum interference in the field ionization of {Rydberg} atoms},
    volume = {92},
    url = {https://link.aps.org/doi/10.1103/PhysRevA.92.043412},
    doi = {10.1103/PhysRevA.92.043412},
    number = {4},
    urldate = {2026-08-17},
    journal = {Physical Review A},
    publisher = {American Physical Society},
    author = {Feynman, Rachel and Hollingsworth, Jacob and Vennettilli, Michael and Budner, Tamas and Zmiewski, Ryan and Fahey, Donald P. and Carroll, Thomas J. and Noel, Michael W.},
    month = oct,
    year = {2015},
    pages = {043412},
}

@article{mcculloch_field_2017,
    title = {Field ionization of {Rydberg} atoms for high-brightness electron and ion beams},
    volume = {95},
    url = {https://link.aps.org/doi/10.1103/PhysRevA.95.063845},
    doi = {10.1103/PhysRevA.95.063845},
    number = {6},
    urldate = {2026-08-17},
    journal = {Physical Review A},
    publisher = {American Physical Society},
    author = {McCulloch, A. J. and Speirs, R. W. and Grimmel, J. and Sparkes, B. M. and Comparat, D. and Scholten, R. E.},
    month = jun,
    year = {2017},
    pages = {063845},
}

@article{chen_electric_2025,
    title = {Electric field sensing via {Rydberg} electromagnetically induced transparency using {Zeeman} and {Stark} effects},
    volume = {138},
    issn = {0021-8979},
    url = {https://doi.org/10.1063/5.0277790},
    doi = {10.1063/5.0277790},
    number = {9},
    urldate = {2026-02-02},
    journal = {Journal of Applied Physics},
    author = {Chen, Yu-Chi and Fang, Shao-Cheng and Su, Hsuan-Jui and Chen, Yi-Hsin},
    month = sep,
    year = {2025},
    pages = {094501},
}

@article{holloway_electromagnetically_2022,
    title = {Electromagnetically induced transparency based {Rydberg}-atom sensor for traceable voltage measurements},
    volume = {4},
    issn = {2639-0213},
    url = {https://pubs.aip.org/aqs/article/4/3/034401/2835268/Electromagnetically-induced-transparency-based},
    doi = {10.1116/5.0097746},
    number = {3},
    urldate = {2024-09-11},
    journal = {AVS Quantum Science},
    author = {Holloway, Christopher L. and Prajapati, Nikunjkumar and Sherman, Jeffery A. and Rüfenacht, Alain and Artusio-Glimpse, Alexandra B. and Simons, Matthew T. and Robinson, Amy K. and La Mantia, David S. and Norrgard, Eric B.},
    month = sep,
    year = {2022},
    pages = {034401},
}

@article{li_super_2023,
    title = {Super low-frequency electric field measurement based on {Rydberg} atoms},
    volume = {31},
    copyright = {© 2023 Optica Publishing Group},
    issn = {1094-4087},
    url = {https://opg.optica.org/oe/abstract.cfm?uri=oe-31-18-29228},
    doi = {10.1364/OE.499244},
    number = {18},
    urldate = {2026-03-30},
    journal = {Optics Express},
    publisher = {Optica Publishing Group},
    author = {Li, Ling and Jiao, Yuechun and Hu, Jinlian and Li, Huaqiang and Shi, Meng and Zhao, Jianming and Jia, Suotang},
    month = aug,
    year = {2023},
    pages = {29228--29234},
}

@article{noaman_rydberg-atom_2023,
    title = {Rydberg-{Atom} {Sensors} in {Bichromatic} {Radio}-{Frequency} {Fields}},
    volume = {20},
    issn = {2331-7019},
    url = {https://link.aps.org/doi/10.1103/PhysRevApplied.20.024068},
    doi = {10.1103/PhysRevApplied.20.024068},
    number = {2},
    urldate = {2024-09-11},
    journal = {Physical Review Applied},
    author = {Noaman, Mohammad and Booth, Donald W. and Shaffer, James P.},
    month = aug,
    year = {2023},
    pages = {024068},
}

@article{ravets_coherent_2014,
    title = {Coherent dipole–dipole coupling between two single {Rydberg} atoms at an electrically-tuned {Förster} resonance},
    volume = {10},
    issn = {1745-2473, 1745-2481},
    url = {https://www.nature.com/articles/nphys3119},
    doi = {10.1038/nphys3119},
    number = {12},
    urldate = {2026-08-18},
    journal = {Nature Physics},
    author = {Ravets, Sylvain and Labuhn, Henning and Barredo, Daniel and Béguin, Lucas and Lahaye, Thierry and Browaeys, Antoine},
    month = dec,
    year = {2014},
    pages = {914--917},
}

@article{wang_atom-interferometric_2015,
    title = {Atom-interferometric measurement of {Stark} level splittings},
    volume = {92},
    url = {https://link.aps.org/doi/10.1103/PhysRevA.92.033619},
    doi = {10.1103/PhysRevA.92.033619},
    number = {3},
    urldate = {2026-08-18},
    journal = {Physical Review A},
    publisher = {American Physical Society},
    author = {Wang, Limei and Zhang, Hao and Zhang, Linjie and Raithel, Georg and Zhao, Jianming and Jia, Suotang},
    month = sep,
    year = {2015},
    pages = {033619},
}

@article{jiao_spectroscopy_2016,
    title = {Spectroscopy of cesium {Rydberg} atoms in strong radio-frequency fields},
    volume = {94},
    url = {https://link.aps.org/doi/10.1103/PhysRevA.94.023832},
    doi = {10.1103/PhysRevA.94.023832},
    number = {2},
    urldate = {2026-08-18},
    journal = {Physical Review A},
    publisher = {American Physical Society},
    author = {Jiao, Yuechun and Han, Xiaoxuan and Yang, Zhiwei and Li, Jingkui and Raithel, Georg and Zhao, Jianming and Jia, Suotang},
    month = aug,
    year = {2016},
    pages = {023832},
}

@article{kitching_atom-based_2025,
    title = {Atom-based quantum sensing of electromagnetic fields},
    volume = {12},
    copyright = {© 2025 Optica Publishing Group},
    issn = {2334-2536},
    url = {https://opg.optica.org/optica/abstract.cfm?uri=optica-12-12-2008},
    doi = {10.1364/OPTICA.569334},
    number = {12},
    urldate = {2026-08-18},
    journal = {Optica},
    publisher = {Optica Publishing Group},
    author = {Kitching, John and Shaffer, James P. and Budker, Dmitry},
    month = dec,
    year = {2025},
    pages = {2008--2022},
}

@article{schlossberger_rydberg_2024,
    title = {Rydberg states of alkali atoms in atomic vapour as {SI}-traceable field probes and communications receivers},
    copyright = {2024 This is a U.S. Government work and not under copyright protection in the US; foreign copyright protection may apply},
    issn = {2522-5820},
    url = {https://www.nature.com/articles/s42254-024-00756-7},
    doi = {10.1038/s42254-024-00756-7},
    urldate = {2024-09-16},
    journal = {Nature Reviews Physics},
    publisher = {Nature Publishing Group},
    author = {Schlossberger, Noah and Prajapati, Nikunjkumar and Berweger, Samuel and Rotunno, Andrew P. and Artusio-Glimpse, Alexandra B. and Simons, Matthew T. and Sheikh, Abrar A. and Norrgard, Eric B. and Eckel, Stephen P. and Holloway, Christopher L.},
    month = sep,
    year = {2024},
    pages = {1--15},
}

@article{manchaiah_frequency_2026,
    title = {Frequency comb behavior of time crystals in an rf-driven dissipative {Rydberg} system},
    volume = {8},
    url = {https://link.aps.org/doi/10.1103/fg8f-q4rs},
    doi = {10.1103/fg8f-q4rs},
    number = {3},
    urldate = {2026-08-18},
    journal = {Physical Review Research},
    publisher = {American Physical Society},
    author = {Manchaiah, Dixith and Watterson, William J. and Holloway, Christopher L.},
    month = jul,
    year = {2026},
    pages = {033115},
}

@article{norrgard_quantum_2021,
    title = {Quantum blackbody thermometry},
    volume = {23},
    issn = {1367-2630},
    url = {https://iopscience.iop.org/article/10.1088/1367-2630/abe8f5},
    doi = {10.1088/1367-2630/abe8f5},
    number = {3},
    urldate = {2024-09-11},
    journal = {New Journal of Physics},
    author = {Norrgard, Eric B and Eckel, Stephen P and Holloway, Christopher L and Shirley, Eric L},
    month = mar,
    year = {2021},
    pages = {033037},
}

@article{schlossberger_primary_2025,
    title = {Primary quantum thermometry of mm-wave blackbody radiation via induced state transfer in {Rydberg} states of cold atoms},
    volume = {7},
    issn = {2643-1564},
    url = {https://link.aps.org/doi/10.1103/PhysRevResearch.7.L012020},
    doi = {10.1103/PhysRevResearch.7.L012020},
    number = {1},
    urldate = {2026-08-18},
    journal = {Physical Review Research},
    author = {Schlossberger, Noah and Rotunno, Andrew P. and Eckel, Stephen P. and Norrgard, Eric B. and Manchaiah, Dixith and Prajapati, Nikunjkumar and Artusio-Glimpse, Alexandra B. and Berweger, Samuel and Simons, Matthew T. and Shylla, Dangka and Watterson, William J. and Patrick, Charles and Meraki, Adil and Talashila, Rajavardhan and Younes, Amanda and La Mantia, David S. and Holloway, Christopher L.},
    month = jan,
    year = {2025},
    pages = {L012020},
}

@misc{prajapati_data_nodate,
    title = {Data for "{Floquet} {Interpretation} of {Avoided} {Crossings} in {AC} {Stark}-{Shifted} {Rydberg}-{EIT} {Spectra}"},
    url = {doi:10.18434/mds2-4297},
    urldate = {2026-09-03},
    author = {Prajapati, Nikunjkumar},
}

\begin{hidden}
\section{Notes}

\begin{itemize}

\item  It provides the AC Stark shifts resulting from the coupling between a large manifold of Rydberg states through a single rf field

    \item Avoided crossings is characterized not only by a splitting of the energies but also by a redistribution of oscillator strength. Measuring both the energy splitting and the relative peak amplitudes provides more information than the energies alone: The gap reveals the coupling strength, while the intensity exchange reveals how the underlying quantum states are mixing 
    \item The bare state character is exchanged at the avoided crossing. This leads to transfer of oscillator strength and change of EIT peak amplitudes.
    \item Shirley method also gives the Stark shifting. Include these points in the manuscript.
    \item Need to include the n=60, 400 MHz result also and explain why it doesn't have avoided crossings.
   
    \item Shirley method captures the effects of repeated absorption and stimulated emission of rf quanta through the coupling between different q-sectors, without ever explicitly introducing photon-number states.

    \item Shirley method successfully predicts the one and multi-photon resonances, avoided crossings, AC Stark Shifts, redistribution of the oscillator strength and complex low-frequency spectra.

    \item The rf source is a microwave generator producing an enormous coherent EM field. Such a field contains an extremely large average photon number, so treating it classically is an excellent approximation.
    \item The avoided crossings are between the physical dressed states, as a result of photon-assisted hybridization.
    
    \item At an avoided crossing, the optical strength is exchanged between the interacting branches because the dominant bare states character of the eigen vectors swaps across the crossing. The exchange of eigen vector composition is often a clear signature of state mixing than the energy split alone.

    \item The off-diagonal terms contain all rf couplings.
    
    \item Obtain the multi-photon coupling graphs and include them in the manuscript. The graphs should connect the initial and final states. Include the character of the bare state in the avoided crossings diagrams.
    
    \item How are transition probabilities related to the avoided crossings?
    
      \item Floquet eigenstates are the physical dressed atom-field states that correspond to the classical-field limit, to the dressed states obtained from a fully quantized atom-plus-field Hamiltonian.

       \item Shirley has shown that his Floquet states can be interpreted physically as quantum field states. His Floquet quasi-energy diagram is identical to the dressed-atom picture.
\end{itemize}

\end{hidden}

\end{document}